\newcommand*{\detset}[1]{{S}^{(#1)}}
\newcommand*{\invEh}{$E_{\rm h}^{-1}$\xspace}
\newcommand{\trial}[0]{\Omega}
\newcommand{\proj}[0]{\hat{P}_0}
\newcommand{\expTaylor}[0]{\rm expTaylor}
\newcommand{\expCheby}[0]{\rm expCh}
\newcommand{\wallCheby}[0]{\rm wallCh}
\newcommand{\e}[1]{\ensuremath{\times 10^{#1}}}

\newif\ifpreprint

\preprintfalse % Enable for double column reprint

\ifpreprint
\documentclass[aip,jcp,amsmath,amssymb,preprint]{revtex4-1}
\else
\documentclass[aip,jcp,amsmath,amssymb,reprint]{revtex4-1}
\fi

\usepackage[english]{babel}

\usepackage{graphicx}% Include figure files
\usepackage{dcolumn}% Align table columns on decimal point
\usepackage{bm}% bold math
\usepackage{amsfonts}
\usepackage{pbox}
\usepackage{braket}
\usepackage{multirow}
\usepackage{threeparttable}
\usepackage{float}
\usepackage{xspace}
\usepackage{color}
\usepackage{fancyvrb}
\usepackage{adjustbox}
\usepackage[colorlinks = true,
            linkcolor = blue,
            urlcolor  = black,
            citecolor = blue,
            anchorcolor = black]{hyperref}
\usepackage{booktabs}
\usepackage{tabularx}

\newcommand*{\kcal}{kcal mol$^{-1}$\xspace}

\newcommand*{\Eh}{$E_{\rm h}$\xspace}

\newcommand{\cop}[1]{\hat{a}^\dag_{#1}}
\newcommand{\aop}[1]{\hat{a}_{#1}}

\newcommand{\methodname}[0]{{DSRG-MRPT2}\xspace}

\begin{document}
%\doublespacing

%
% Title
%
\title{A deterministic projector configuration interaction approach for the ground state of quantum many-body systems}
%strongly correlated electrons}

%
% Authors
%
\author{Tianyuan Zhang}
\author{Francesco A. Evangelista}
\email{francesco.evangelista@emory.edu}
\affiliation{Department of Chemistry and Cherry L. Emerson Center for Scientific Computation, Emory University, Atlanta, Georgia 30322, USA}

%
% Abstract
%
\begin{abstract}
{\footnotesize In this work we propose a novel approach to solve the Schr\"{o}dinger equation which combines projection onto the ground state with a path-filtering truncation scheme.
The resulting projector configuration interaction (PCI) approach realizes a deterministic version of the full configuration interaction quantum Monte Carlo (FCIQMC) method \lbrack Booth, G. H.; Thom, A. J. W.; Alavi, A. \textit{J. Chem. Phys.} \textbf{2009}, \textit{131}, 054106\rbrack.
To improve upon the linearized imaginary-time propagator, we develop an optimal projector scheme based on an exponential Chebyshev expansion in the limit of an infinite imaginary time step.
After writing the exact projector as a path integral in determinant space, we introduce a path filtering procedure that truncates the size of the determinantal basis and approximates the Hamiltonian.
The path filtering procedure is controlled by one real threshold that determines the accuracy of the PCI energy and is not biased towards any determinant.  Therefore, the PCI approach can equally well describe static and dynamic electron correlation.
This point is illustrated in benchmark computation on N$_2$  at both equilibrium and stretched geometries.  In both cases, the PCI achieves chemical accuracy with wave functions that contain less than 0.5\% of the full CI space.
We also report computations on the ground state of C$_2$ with up to quaduple-$\zeta$ basis sets and wave functions as large as 200 million determinants, which allow a direct comparison of the  PCI, FCIQMC, and density matrix renormalization group (DMRG) methods.
The size of the PCI wave function grows modestly with the number of unoccupied orbitals and its accuracy may be tuned to match that of FCIQMC and DMRG.}
\end{abstract}

\maketitle

%
% Introduction
%
\section{Introduction}
The full configuration interaction (FCI) approach provides the exact solution to the electronic Schr\"{o}dinger equation within a finite one-particle basis set.\cite{sherrillCIrev}
However, since the number of FCI wave function parameters grows rapidly with system size, this approach is only feasible for few electrons distributed in a small number of orbitals.\cite{rossi1999full}
Contrary to what is suggested by this observation, a large body of evidence has been amassed that shows that the information content of \textit{molecular} wave functions is just a small fraction of the size of the FCI basis.\cite{knowles1989unlimited}
For example, for wave functions dominated by one Slater determinant, truncated coupled cluster theory can recover a large fraction of the dynamical correlation energy at a cost that is polynomial in the number of electrons.\cite{bartlett2007coupled}
However, in the case of strongly correlated electrons, the problem of finding general polynomial-scaling wave function methods is still open.\cite{bartlett1981many,dagotto2005complexity}

Several strategies have been suggested to overcome the exponential cost of FCI and FCI performed in a complete active space (CASCI), including selected CI approaches that truncate FCI space,\cite{buenker1974individualized,buenker1975energy,huron1973iterative, evangelisti1983convergence, meller1994size,bender1969studies,langhoff1973configuration,angeli1997multireference, angeli1997multireference2,angeli1998multireference,Olsen:1988ut,Ivanic:2003tx,Ma:2011hz,LiManni:2013jm,schriber2016communication}
tensor factorization,\cite{white1992density,white1999ab,chan2002highly,Chan2004State,kurashige2009high, mizukami2012more,kurashige2014theoretical,kurashige2013entangled,Booth2012Excited,Olivares-Amaya2015ab,Vidal:2007kx,Nakatani:2013hw,Nakano:2000uw,parker2013communication,Parker:2014fg,Murg:2015vg,Szalay:2015fj,Bohm:2016by}
alternative configuration interaction and coupled cluster methods,\cite{Mayhall:2014eu,Small:2014ed,Stein:2014hi,Bulik:2015gk}
symmetry breaking and restoration,\cite{Tsuchimochi:2009cl,JimenezHoyos:2012gx,RodriguezGuzman:2013bw}
and Monte-Carlo methods.\cite{Greer1995Estimating,coe2014applying,  coe2013state, coe2012development,Gyorffy2008Monte,sugiyama1986auxiliary,honma1995diagonalization,al2006auxiliary,ohtsuka2008projector,shi2013symmetry,Booth2009Fermion,Cleland2010Survival,Booth2011Breaking,Cleland2011Study,Cleland2012Taming,Thomas2014Symmetry,Booth2014Linear}
Recently, Monte-Carlo methods that stochastically sample the wave function in the space of Slater determinants have received wide attention.
The Monte-Carlo CI method (MCCI) uses stochastic sampling to find an optimal space of orthogonal Slater determinants.\cite{Greer1995Estimating,coe2014applying,  coe2013state, coe2012development,Gyorffy2008Monte}
MCCI may be viewed as a stochastic version of selected CI since at each iteration the energy is obtained by diagonalizing the Hamiltonian is a subset of the FCI space.\cite{buenker1974individualized,buenker1975energy,huron1973iterative, evangelisti1983convergence}
Another stochastic method is the auxiliary-field QMC (AFQMC) approach.\cite{sugiyama1986auxiliary,honma1995diagonalization,al2006auxiliary,shi2013symmetry} AFQMC uses a projector formalism and differs from MCCI in its use of non-orthogonal Slater determinants and the fact that the wave function is sampled stochastically. 
Deterministic analogs of the AFQMC approach have also been developed, including the path-integral renormalization group method\cite{imada2000path,imai2007applications} and the non-orthogonal multicomponent adaptive greedy iterative compression approach of McClean and Aspuru-Guzik.\cite{mcclean2015compact}
%In this work the authors combine imaginary-time propagation with compressed sensing to find a compact representation of the wave function.

An alternative to the MCCI and AFQMC methods is the FCI Quantum Monte-Carlo (FCIQMC) method developed by Alavi and co-workers.\cite{Booth2009Fermion,Cleland2010Survival,Booth2011Breaking,Cleland2011Study,Cleland2012Taming,Thomas2014Symmetry,Booth2014Linear} FCIQMC is a projector Monte-Carlo method that samples the imaginary-time propagator in a space of orthogonal Slater determinants.
%FCIQMC obtains the ground state energy via stochastic sampling of the imaginary-time propagator.  
By working in a basis of Slater determinants, FCIQMC can more easily account for the annihilation of walkers of different sign.  This feature ameliorates the sign problem, but a large number of walkers are necessary to accurately sample the FCI space of determinants.
The initiator approximation\cite{Cleland2010Survival} reduces the number of walkers required in FCIQMC and increases the sign coherence of the sampling.
Furthermore, a semi-stochastic version of FCIQMC (SFCIQMC) was later introduced,\cite{petruzielo2012semistochastic, blunt2015semi-stochastic,umrigar2015observations} which shows that treating part of the imaginary-time projection deterministically accelerates convergence and reduces statistical uncertainty.

The improvements to the performance of FCIQMC brought by treating part of determinant space deterministically raises the interesting question of whether a fully deterministic projector method might be even more advantageous.
As pointed out by Tubman and co-workers,\cite{tubman2016deterministic} the stochastic dynamics of FCIQMC reinterpreted in a deterministic way corresponds to a truncation criterion for selected CI.
In this work, we demonstrate an alternative route to create a deterministic analog of FCIQMC.
An important feature of our new method is the 
use of a projection scheme that simultaneously selects an optimal CI space and approximately diagonalizes the Hamiltonian.
The resulting computational method is named projector configuration interaction (PCI).
The PCI approach automatically identifies the most important determinants that contribute to the ground state wave function, therefore, it can treat both dynamic and static electron correlation.

The PCI methods presents two major differences with respect to FCIQMC.
As in other projector Monte-Carlo methods, FCIQMC relies on a linearized approximation to the imaginary-time projector obtained by Taylor expansion.
One of the major drawbacks of this approximation is that a small time step is required to guarantee convergence to the ground state, the length of which is bound by the inverse spectral radius of the Hamiltonian.
Following the work of Kosloff and Tal-Ezer\cite{kosloff1986direct}, we overcome this limitation by using a Chebyshev expansion of the exponential projector.\cite{zhu1994orthogonal,kouri1995acceleration,parker1996matrix,Chen1999chebyshev,boyd2001chebyshev}
%To overcome this limitation, in the PCI we employ a Chebyshev expansion of the exponential projector.\cite{kosloff1986direct,zhu1994orthogonal,kouri1995acceleration,parker1996matrix,Chen1999chebyshev,boyd2001chebyshev}
%Chebyshev polynomials have been widely used in quantum chemistry study for eigenstates.\cite{Chen1999chebyshev} For example, filter diagonalization (FD) uses Chebyshev polynomials to fit a delta function to filter target states for diagonalization with standard diagonalization methods. Kosloff and Tal-Ezer\cite{kosloff1986direct} also introduced the direct relaxation method for eigenpairs, which successively applys the Chebyshev fitted imaginary-time projector onto a trial wave function.
In particular, we consider the \textit{wall}-Chebyshev projector, which is derived from the Chebyshev representation of the imaginary-time propagator in the limit of an infinite time step.
In this respect, our goal is analogous to that of the $t$ expansion method, in which the $t \rightarrow\infty$ limit of the imaginary-time propagator is expressed using Pad\'{e} approximants.\cite{horn1984t}
The \textit{wall}-Chebyshev generator is shown to be equivalent to a power method with alternating shifts, and it is more efficient than the corresponding Taylor and Chebyshev expansions of the exponential projector.
We also address the issue of replacing Monte-Carlo sampling with a deterministic truncation of the determinant space.
Since projection onto the ground state may be viewed as a path-integral scheme, we apply the idea of path filtering\cite{sim1997filtered,sim1997path,makri1999time,lambert2012quantum} in order to truncate CI space and control accuracy.
In the PCI, path filtering is applied to screen excited determinants generated by projection onto the ground state.
Path filtering is controlled by one threshold parameter, and as a consequence, the PCI forms a family of one-parameter theories that are systematically improvable and equivalent to FCI when path filtering is suppressed.

The paper is organized in the following way.
In section 2, we introduce the formalism of ground state projection, Chebyshev fitting of the imaginary-time propagator, and path filtering.  Section 3 details the PCI algorithm and our implementation and analyzes the sources of error in the PCI approach. 
In section 4 we demonstrate the ability of PCI to adapt to various regimes of electron correlation by applying it to the dissociation of N$_2$.
In the same section, we study the scaling of the PCI cost with respect to basis set size and the size consistency error introduced by the path-filtering approximation.  

\section{Theory}
\subsection{General formalism of ground state projection}
\label{subsec:pfci}

Given the Hamiltonian operator $\hat{H}$, we write its eigenvalues and eigenfunctions as $E_i$ and $\Psi_i$, respectively.
Within a finite computational basis, the Hamiltonian is assumed to have $N$ eigenfunctions, and its spectral radius ($R$) is defined as the difference between the largest ($E_{N-1}$) and smallest ($E_0$) eigenvalues divided by two:
\begin{equation}
R=\frac{E_{N-1}-E_0}{2}
\end{equation}
The goal of projector CI (PCI) is to obtain the ground state wave function $\Psi_0$ starting from a trial wave function $\trial$ via a projector operator $\proj$:
\begin{equation} \label{eq:projection}
\ket{\Psi_0} = N_P \proj \ket{\trial}
\end{equation}
The only assumption concerning the trial wave function is that its overlap with the exact ground state is not zero, that is $\braket{\trial|\Psi_0} \neq 0$.
In Eq.~\eqref{eq:projection}, $N_P$ is a normalization factor introduced to guarantee that $\braket{\Psi_0|\Psi_0} =1$ and the projector operator $\proj$ is assumed to be idempotent ($\proj^2 = \proj$).

We restrict our discussion to a class of projectors that can be written as the infinite product:
\begin{equation}
\proj = \lim_{n \rightarrow \infty} g^n(\hat{H})
\end{equation}
where $g(\cdot)$ is the \textit{generator of the projector} $\proj$ (also abbreviated as \textit{generator} in the following).
The projector generator is assumed to be a real function $g:\mathbb{R}\rightarrow \mathbb{R}$ extended to the domain of Hermitian operators.
Given a generic state vector $\ket{\trial}$, it may be decomposed as a sum over the eigenfunctions of the Hamiltonian as:
\begin{equation}
\ket{\Omega}=\sum_{i} c_i \ket{\Psi_i}
\end{equation}
so that the action of the projector generator $g(\hat{H})$ onto $\ket{\Omega}$ may be written out:
\begin{equation}
g(\hat{H})\ket{\Omega}=\sum_{i}c_ig(\hat{H})\ket{\Psi_i}=\sum_{i}g(E_i) c_i\ket{\Psi_i}
\end{equation}
Thus, the application of a generator onto a trial state vector leads to a new state vector in which the coefficient that multiplies each $\ket{\Psi_i}$ is amplified by a factor  $g(E_i)$, where $E_i$ is the eigenvalue corresponding to $\ket{\Psi_i}$.

For an appropriately chosen generator, the repeated application of $g(\hat{H})$ may be used to amplify the coefficient of the ground state wave function and reduce that of excited states.
A necessary condition for the generator to project a state onto $\Psi_0$ is to satisfy the inequality:
\begin{equation} \label{eq:necessary}
%|g(E_0)| > |g(x)| \text{ for } \forall x \in (E_0, E_{N-1}]
%|g(E_0)| > |g(x)| \quad \forall x > E_0
|g(E_0)| > |g(x)| \quad \forall x \in (E_0, E_{N-1}]
\end{equation}
so that the relative weight of the excited states is reduced by a factor $q_i = g(E_i)/g(E_0)$:
\begin{equation} \label{eq:g_trial}
g(\hat{H})\ket{\Omega}= c_0 \ket{\Psi_0} + \sum_{i = 1}^{N-1}q_i c_i\ket{\Psi_i} \quad |q_i| < 1
\end{equation}
where without loss of generality, we have assumed that $g(x)$ is scaled so that $g(E_0) = 1$.
In practical applications, the range of $\hat{H}$ is unknown, but as discussed in section~\ref{sec:range}, one may obtain upper bounds of $E_0$ and $E_{N-1}$ (here denoted $\tilde{E}_0$ and $\tilde{E}_{N-1}$).
In this case, it is convenient to work with generators that decrease monotonically in the left-neighborhood of $\tilde{E}_0$, that is for any two points $x,y \in [E_0, \tilde{E}_0]$:
\begin{equation} \label{eq:local_monotonicity}
|g(x)| > |g(y)| \text{ if } x < y
\end{equation}
When the monotonicity condition expressed by Eq.~\eqref{eq:local_monotonicity} is satisfied, the projector is guaranteed to converge to the ground state even if $E_0$ and $E_{N-1}$ are approximated with their respective upper bound estimates.
Therefore, in the following discussion we do not distinguish $\tilde{E}_0$ from $E_0$.

\subsection{Rate of convergence of generators.}
The repeated application of the generator onto a trial wave function $\trial^{(0)}$ generates a sequence of vectors:
\begin{equation}
\ket{\trial^{(n)}}=g^n (\hat{H})\ket{\trial^{(0)}}
\end{equation}
which in the limit of $n$ that goes to infinity converges to the exact ground state:
\begin{equation}
\ket{\Psi_0} = \lim_{n\rightarrow \infty}\ket{\trial^{(n)}}
\end{equation}
The asymptotic rate of convergence of this sequence is defined as:
\begin{equation}
\label{eq:convergence_factor_def}
\mu = \lim_{n\rightarrow\infty}\frac{\|\trial^{(n+1)} - \Psi_0\|}{\|\trial^{(n)} - \Psi_0\|}  = \max_i |q_i|
\end{equation}
and is given by the $q_i$ factor with the largest absolute value.

When the rate of convergence is controlled by the first excited state, that is $\mu = |q_1|$, and the energy difference $E_1 - E_0$ is small compared to the spectral radius, then we can approximate $\mu$ as:
\begin{equation}
\label{eq:convergence_factor_approx}
\mu=\left|\frac{g(E_1)}{g(E_0)}\right|\approx \left|1+ g'(E_0) \cdot(E_1 - E_0)\right|
\end{equation}
where $g'(E_0)$ is the first derivative of $g(x)$ at $E_0$.
%Since $(E_1 - E_0)$ is a small constant for a specific system, the rate of convergence depends only on $g'(\lambda_1)/g(\lambda_{1})$.
Hence, we can define the convergence factor $\gamma$ for $g(x)$ as
\begin{equation} \label{eq:convergence_factor}
\gamma=- g'(E_0)
\end{equation}
It is possible to show that the number of times one must apply $g(\hat{H})$ to a trial wave function in order to achieve a certain level of accuracy is inversely proportional to $\gamma$.
Therefore, the convergence factor provides a quantitative estimate of the numerical efficiency of a generator.
Generators with large convergence factors are in general preferable as they are expected to reduce the computational cost of the PCI.
The parameters that enter the definition of all the generators discussed in this work and their corresponding convergence factor are summarized in Table~\ref{tab:generators}.

%
% Table 1: Projector Generators
%
\begin{table*}[t!]
\footnotesize
\begin{threeparttable}
  \caption{Comparison of different projector generators.  The form of  the projector generator [$g(x)$] and convergence factor ($\gamma$) is given as a function of the time step ($\tau$), the spectral radius of the Hamiltonian ($R$), and the order of the polynomial expansion ($m$). }
   \label{tab:generators}
    { 
{\def\arraystretch{2.5}\tabcolsep=0pt    
    
     \begin{tabular*}{6.5in}{@{\extracolsep{\fill} }llll}
       \hline
       
       \hline
       Generator & Parameters &  $g(x)$ & Convergence factor ($\gamma$)\\
       \hline
       Exponential & $\tau$ & $e^{-\tau (x-E_0)}$ & $\tau$ \\
       Linear & $\tau$ & $1-\tau(x-E_0)$  & $\tau<\frac{1}{R}$\tnote{*}\\
%       Delta &  & $\delta(\det(x-E_0\hat{I}))$ & $+\infty$\\
       Exp-Taylor & $\tau$,$m$ & $\displaystyle\sum_{k=0}^m \frac{1}{k!}(-\tau)^k(x-E_0)^k$ & $\tau<\frac{m+1}{2R}$\tnote{**}\\
       Exp-Chebyshev & $\tau$,$R$,$m$ & $\displaystyle C_{m}(\tau R)  \sum_{k=0}^m(2-\delta_{k0})I_k(\tau R)T_k\left(-\frac{x-E_0-R}{R}\right)$ & $\displaystyle\frac{\sum_{k=1}^m 2I_k(\tau R)k^2}{R\sum_{k=0}^m(2-\delta_{k0})I_k(\tau R)}<\frac{m(m+1)}{3R}$\\
       Wall-Chebyshev & $R$,$m$ & $\displaystyle\frac{1}{2m+1}\sum_{k=0}^m(2-\delta_{k0})T_k\left(-\frac{x-E_0-R}{R}\right)$ & $\displaystyle\frac{m(m+1)}{3R}$\\
       \hline
       
       \hline
      \end{tabular*}
      }
    }
    \begin{tablenotes}
    \footnotesize
    \item[*] In order to converge onto the ground state wave function, the time step must satisfy the condition: $\tau<\frac{1}{R}$.
    \item[**] In order to converge onto the ground state wave function, the time step must satisfy the condition: $|\sum_{k=0}^m \frac{1}{k!}(-\tau)^k(2R)^k|<1$.  From this expression one may derive the upper bound: $\tau<\frac{m+1}{2R}$.
    \end{tablenotes}
\end{threeparttable}
\end{table*}

\subsection{Taylor and Chebyshev expansions of the imaginary-time propagator}
The projector generator corresponding to the imaginary-time propagator, $\lim_{\beta \rightarrow \infty} e^{-\beta (\hat{H}-E_0)}$, is the exponential generator ($g_{\rm exp}$), defined as:
\begin{equation} \label{eq:exp_gen}
g_{\rm exp}(x)=e^{-\tau (x - E_0)}
\end{equation}
This generator satisfies both conditions Eqs.~\eqref{eq:necessary} and \eqref{eq:local_monotonicity}.
Nevertheless, it is not expressed as a polynomial of the Hamiltonian and therefore, to make its evaluation computationally viable it must be approximated with a polynomial expansion.
To evaluate the projector based on the exponential generator [Eq.~\eqref{eq:exp_gen}] it is necessary to expand $g_{\rm exp}(x)$ into a polynomial series.

An $m$-th order Taylor expansion of $g_{\rm exp}(x)$ centered around $E_0$ yields the generator:
\begin{equation}
g_{\expTaylor}(x)
= \sum_{k=0}^m \frac{1}{k!}(-\tau)^k (x-E_0)^k
\end{equation}
which has convergence factor $\gamma_{\expTaylor} = \tau$ independent of the truncation order $m$.
Consequently, there is not gain in efficiency when $g_{\expTaylor}(x)$ is expanded beyond $m=1$.
More importantly, the Taylor expansion is only accurate near $E_0$, and since the error grows as a power of $\tau(x-E_0)$, a very small value of $\tau$ may be required to satisfy the necessary condition for the convergence of the projector [see Eq.~\eqref{eq:necessary}].

Note, that the first-order Taylor expansion of the exponential:
\begin{equation}
g_{\rm linear}(x)=1-\tau (x - E_0) =
 - \tau(x-s)
% - \tau(x-(E_0+\frac{1}{\tau}))
\end{equation}
 is equivalent to a power method with shift $s=E_0+\frac{1}{\tau}$.
 In order to converge to the ground state wave function, the shift must be chosen to satisfy $s>R$.
The corresponding convergence factor is bound by the inverse of the spectral range of the Hamiltonian:
\begin{equation} \label{eq:gamma_lin}
\gamma_{\rm linear}= \tau <\frac{1}{R}
\end{equation}

An alternative approximation of the exponential with better error control is an expansion in terms of Chebyshev polynomials (for example, see Refs.~\citenum{kosloff1986direct} and \citenum{boyd2001chebyshev}).
 Following Kosloff and Tal-Ezer,\cite{kosloff1986direct} we write the $m$-th order Chebyshev polynomial fitting of the exponential generator as:
\begin{equation}
\begin{split}
g_{\expCheby}(x)
= C_{m}(\tau R) \sum_{k=0}^m(2-\delta_{k0})I_k(\tau R)T_k\left(-\frac{x-E_0-R}{R}\right)
\end{split}
\end{equation}
where $C_{m}(\tau R)=1/\left(\sum_{k=0}^m(2-\delta_{k0})I_k(\tau R)\right)$ is a scaling factor that guarantees $g_{\expCheby}(E_0) =1$, $\delta_{k0}$ is a Kronecker delta, $I_k$ is the $k$-th modified Bessel function of the first kind, and $T_k$ is the $k$-th order Chebyshev polynomial.

\begin{figure}[!hbp]
    	\includegraphics[width=3.375in]{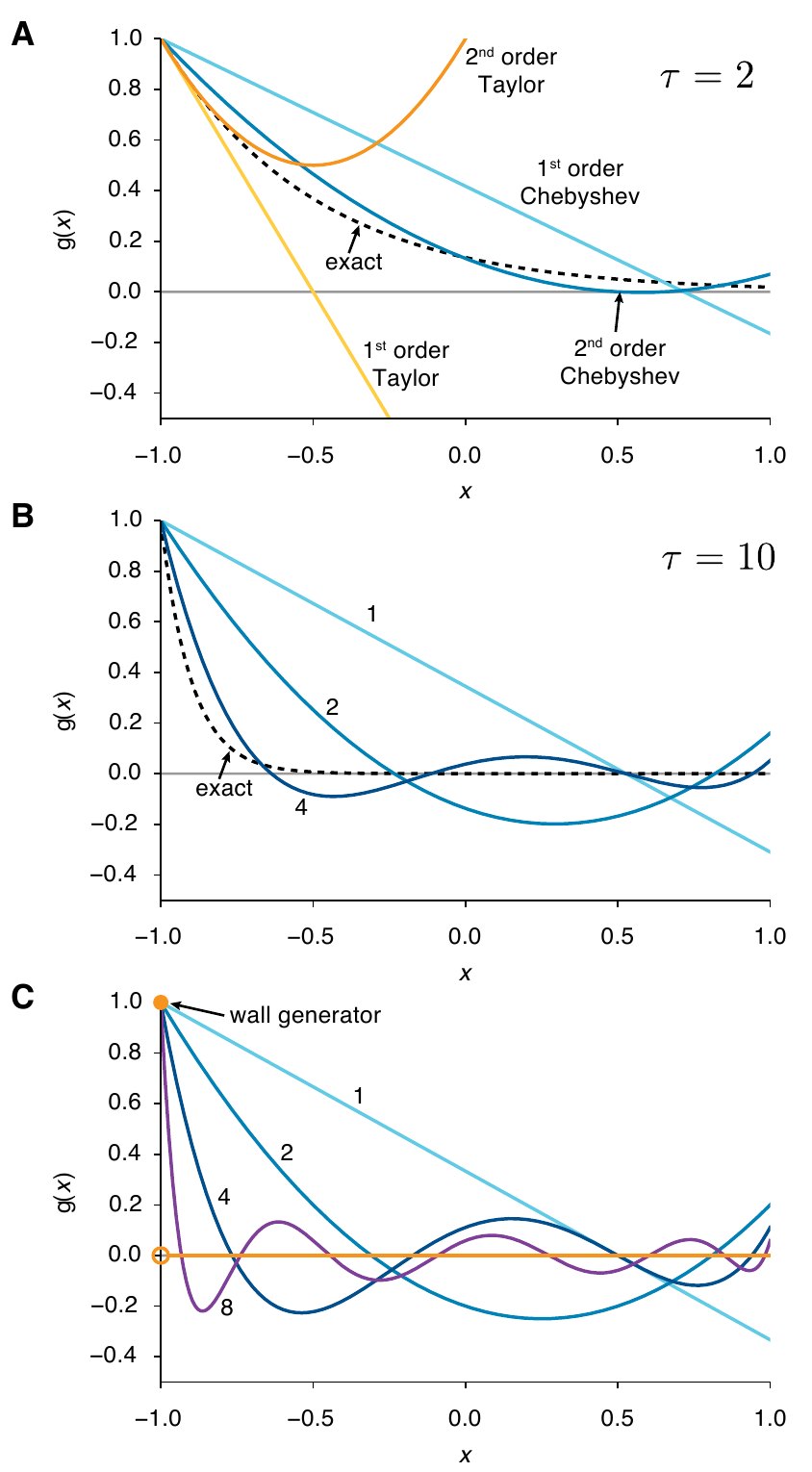}
    	\caption{Polynomial approximations of the exponential generator [Eq.~\eqref{eq:exp_gen}] and the $\tau\rightarrow\infty$ limit of the exponential generator (wall generator) [Eq.~\eqref{eq:wall_generator}] plotted in the range $[-1,1]$ \Eh.
	(A) Taylor and Chebyshev approximation of the exponential generator for $\tau = 2$ \invEh at order 1 and 2.  (B) Chebyshev approximation of the exponential generator for $\tau = 10$ \invEh at order 1, 2, and 4.  (C) Chebyshev approximation of the wall generator at order 1, 2, 4, and 8.}
	\label{fig:exp-wall-taylor-chebyshev}
\end{figure}

%Figure~\ref{fig:exp-wall-taylor-chebyshev}A shows first- and second-order fits of the exponential evaluated for $\tau = 0.1$ and for the range $[-19.11,9.55]$ \Eh , which corresponds to the range of eigenvalues estimated for a test system (BeH$_2$).
%This plot illustrates the points made above: i) the Taylor expansion of the exponential is accurate only near the expansion point (in this case $E_0 = -19.11$ \Eh) and ii) the Chebyshev expansion is well-behaved on the entire range.
%The bottom panel of Figure~\ref{fig:exp-wall-taylor-chebyshev}B shows Chebyshev expansion for the same range but with $\tau = 1$.
%In this case the fitting error is significantly larger and the convergence of the Chebyshev expansion with respect to the order $m$ is slow.
Figure~\ref{fig:exp-wall-taylor-chebyshev}A shows first- and second-order Taylor and Chebyshev expansions of the exponential evaluated for $\tau = 2$ \invEh in the range $[-1,1]$ \Eh.
This plot illustrates the points made above: i) the Taylor expansion of the exponential is accurate only near the expansion point (in this case $E_0 = -1$ \Eh) and ii) the Chebyshev expansion is well behaved on the entire range.
Figure~\ref{fig:exp-wall-taylor-chebyshev}B shows the Chebyshev expansion for the same range but with $\tau = 10$ \invEh.
In this case the fitting error is larger and the convergence of the Chebyshev expansion with respect to the order $m$ is slower than the case $\tau = 2$ \invEh.
Nevertheless, even though the Chebyshev expansion for $\tau = 10$ \invEh does not accurately match the exponential function, it is still a valid projector generator since it satisfies Eqs.~\eqref{eq:necessary} and \eqref{eq:local_monotonicity}.
%This indicates that a Chebyshev expansion up to a certain order cannot fit exponential generator with arbitrarily large $\tau$ accurately. In Figure~\ref{fig:exp-wall-taylor-chebyshev}A and \ref{fig:exp-wall-taylor-chebyshev}B, the first-order Chebyshev fit for $\tau = 10$ \invEh does not differ noticeably from the one for $\tau = 2$ \invEh. There seems to be a limit where a certain order Chebyshev fit remain unchanged when $\tau$ for the exponential is sufficiently large. What's the limit?

\subsection{An improved generator: the wall generator and its Chebyshev expansion.}
In the previous subsection we discussed how to improve the accuracy of the Taylor expansion of the exponential generator via Chebyshev fitting.
Ideally, the best projector generator is the the  \textit{wall} function, defined as: 
\begin{equation}
g_{\rm wall}(x) =
\begin{cases}
0 \text{ for } x > E_0\\
1 \text{ for } x = E_0\\
\infty \text{ for } x < E_0
\end{cases}
\end{equation}
This generator may be viewed as the $\tau\rightarrow\infty$ limit of the exponential generator:
%However, the resulting Chebyshev series has a convergence factor that for small values of $\tau$ is identical to the convergence factor of a Taylor expansion.
%However, the convergence factor of the exponential Chebyshev generator may be further improved by taking the limit $\tau\rightarrow\infty$:
\begin{equation} \label{eq:wall_generator}
g_{\rm wall}(x) = \lim_{\tau\rightarrow\infty} e^{-\tau (x - E_0)}
\end{equation}

Despite the fact that neither definitions of $g_{\rm wall}(x)$ are computationally viable, we can still approximate the wall generator using a Chebyshev expansion, by taking the $\tau\rightarrow\infty$ limit of the $m$-th order exponential Chebyshev generator:
\begin{equation}\label{eq:chebywall}
\begin{split}
g_{\wallCheby}(x)=& \lim_{\tau\rightarrow\infty} g_{\expCheby}(x) \\
=&\frac{1}{2m+1}\sum_{k=0}^m(2-\delta_{k0})T_k\left(-\frac{x-E_0-R}{R}\right)
% \\
%=&\frac{1}{2m+1}D_m\left(\arccos \left(-\frac{x-E_0-R}{R}\right)\right),
\end{split}
\end{equation}
where we used the fact that $\lim_{\tau\rightarrow\infty} I_{k+1}(\tau R)/I_{k}(\tau R) = 1$.\cite{amos1974computation}
Note that this polynomial is a special case of the Chebyshev expansion of the delta distribution with the origin translated to the lower bound of the fitting range.\cite{zhu1994orthogonal,kouri1995acceleration,parker1996matrix,Chen1999chebyshev}

The wall-Chebyshev generators of order 1, 2, 4, and 8 are plotted in Figure~\ref{fig:exp-wall-taylor-chebyshev}C.
An important property of the wall-Chebyshev generator is that for values of $x$ less than $E_0$ these functions are monotonic and diverge when $x \rightarrow -\infty$.
Therefore they satisfy Eq.~\eqref{eq:local_monotonicity} and are able to converge onto the ground state even when the range of $\hat{H}$ is not known precisely.

The Chebyshev expansion of the wall generator may  shown to converge with factor
\begin{equation}
\gamma_{\wallCheby}=\frac{m(m+1)}{3R}
\end{equation}
which is the largest one among all the polynomial generators discussed in this work.
It is important to note that although we can design generators with even larger convergence factors, an efficient generator must also efficiently suppress high energy excited states.
For example, the Chebyshev generator, defined as $g_{\rm Ch}(x)=T_k\left(-\frac{x-E_0-R}{R}\right)$ gives $\gamma_{\rm Ch}=\frac{m^2}{R}$, which is larger than the convergence factor of the generators discussed previously.
However, the convergence of the projector generated by $g_{\rm Ch}(x)$ is slow because the coefficients of high energy excited states are not efficiently reduced.

In each projection generation step, an $m$-th order wall-Chebyshev generator involves the application of the Hamiltonian $m$ times, therefore, it has a cost that is $m$ times that of the linear generator (power method).
Consequently, the theoretical relative acceleration with respect to the most efficient linear generator ($\tau_{\rm linear}=1/R$) is:
\begin{equation}
\frac{\gamma_{\wallCheby}}{m \gamma_{\rm linear}}=\frac{m+1}{3}
\end{equation}
For instance, an $8^{\rm th}$-order wall-Chebyshev generator has a computational cost that is a third of the linear generator with the largest allowed value of $\tau$ (1/$R$).

An important property of the $m$-th order $g_{\wallCheby}(x)$ generator is that it has $m$ distinct real roots in the range $(E_0,E_{N-1})$.
Therefore, it can be decomposed as a product of $m$ linear generators with real shifts:
\begin{equation} \label{eq:wall_decomposition}
g_{\wallCheby}(x) = \prod_{i=1}^m \frac{x-s_i}{E_0-s_i}
\end{equation}
where the shifts $s_i$ are the zeros of $g_{\wallCheby}(x)$.
It is easy to show that the zeros of $g_{\wallCheby}(x)$ can be expressed in closed form as:
\begin{equation} \label{eq:wall_decomposition_zeros}
s_i = E_0 + R \left(1-\cos \frac{i}{m+\frac{1}{2}}\pi \right)
\end{equation}
Eq.~\eqref{eq:wall_decomposition} allows us to implement the wall generator as a product of linear generators applied successively onto a state vector.
Hence, the projector associated with the wall generator may be interpreted as an optimized power method that uses a sequence of energy shifts.
Besides its high efficiency, there are two other advantages of the wall-Chebyshev generator: i) Only two vectors (previous and current) need to be stored during the calculation, in contrast to three vectors necessary for the exp-Chebyshev generator (previous, current and accumulator) and ii) the wall-Chebyshev generator is numerically more stable than the exp-Chebyshev generator since for $\tau \rightarrow \infty$ the numerical evaluation of Bessel functions introduces numerical errors.

\subsection{Determinant selection via path filtering.}
The projector CI discussed in Section~\ref{subsec:pfci} provides an alternative approach to finding the exact ground state.
In this section we show how to combine this methods with path filtering to generate an approach that diagonalizes the Hamiltonian in an optimal subset of FCI space.
We discuss path filtering only for the case of the linear generator and report details for higher-order polynomial generators in appendix~\ref{app:general_path_filtering}.

Consider a normalized trial state $\Omega^{(n)}$ that approximates the exact ground state in the subset $S^{(n)}$ of FCI space:
\begin{equation}
\ket{\Omega^{(n)}} = \sum_{\Phi_J \in S^{(n)}} C_J^{(n)} \ket{\Phi_J}
\end{equation}
where $C_J^{(n)}$ is the coefficient of determinant $\ket{\Phi_J}$ at the $n$-th step.
The action of the linear generator onto $\Omega^{(n)}$ leads to a new state $\tilde{\Omega}^{(n + 1)}$:
\begin{equation}
g_{\rm linear}(\hat{H})\ket{\Omega^{(n)}} = \ket{\tilde{\Omega}^{(n + 1)}}  =\sum_{I} \tilde{C}_I^{(n+1)} \ket{\Phi_I}
\end{equation}
where, in general, the vector of coefficients $\tilde{C}_I^{(n+1)}$ is not normalized.
The coefficients of $\tilde{C}_I^{(n+1)}$ may be expressed as a sum over \textit{spawning amplitudes}, $A_{IJ}^{(n+1)}$:
\begin{equation}
\begin{split}
\tilde{C}_I^{(n+1)} &= \bra{\Phi_I}1-\tau (\hat{H} - E_0) \ket{\Omega^{(n)}}\\
 &= \tau \sum_{\Phi_J \in \detset{n}}   A_{IJ}^{(n+1)}
\end{split}
\end{equation}
where $A_{IJ}^{(n+1)}$ is defined as:
\begin{equation}
\label{eq:spawning_amplitude}
A_{IJ}^{(n+1)} = \frac{1}{\tau}\bra{\Phi_I}1-\tau (\hat{H} - E_0) \ket{\Phi_J} C_J^{(n)}
\end{equation}
The spawning amplitude has the units of a rate and represents the contribution of the $\Phi_J$ component of $\Omega^{(n)}$ that ``flows'' to the coefficient of $\Phi_I$ for state $\tilde{\Omega}^{(n + 1)}$.

The repeated application of the generator onto a trial function generates paths in FCI space that may be filtered (approximated) by thresholding the spawning amplitude.
To this end we introduce a \textit{spawning threshold} $\eta$ and truncate the off-diagonal spawning amplitude as:
\begin{equation}\label{eq:eta_amplitude}
A_{IJ}^{(n)}(\eta) =
\begin{cases}
A_{II}^{(n)} &\text{ if } I = J\\
A_{IJ}^{(n)} \Theta(|A_{IJ}^{(n)}| - \eta) &\text{ if } I \neq J
\end{cases}
\end{equation}
where $\varTheta(x)$ is the Heaviside step function.
Consequently, the PCI update equations for the wave function coefficients are:
\begin{equation}
\label{eq:coef-propagation}
\tilde{C}_I^{(n+1)} = \tau \sum_{\Phi_J \in \detset{n}}   A_{IJ}^{(n+1)}(\eta)
\end{equation}
and the determinant set at step $n+1$ includes only those elements of the FCI space that may be reached from $\detset{n}$ via non-zero amplitudes:
\begin{equation}
\label{eq:detset}
\detset{n+1} = \{ \Phi_I :  \exists \Phi_J \in \detset{n}, A_{IJ}^{(n+1)}(\eta)  \neq 0 \}
\end{equation}
In other words, a determinant is included in $\detset{n+1}$ when there is at least one spawning amplitude that is larger than the spawning threshold.
Note that this selection criterion is analogous to the one used in heat-bath sampling\cite{holmes2016efficient} and accounts both for the weight of a parent determinant, via the factor $C_J^{(n)}$, and for the coupling between parent and spawned determinant, via the matrix element of the linearized generator $\bra{\Phi_I}1-\tau (\hat{H} - \epsilon) \ket{\Phi_J}$.

In order to further reduce the computation cost, the so-called initiator approximation\cite{Cleland2010Survival} is introduced in FCIQMC, which imposes a screening of the determinants that may be spawned.
Translated in the language of the PCI approach, the initiator approximation is equivalent to a path-filtering procedure in which the screening is done according to the absolute value of a determinant coefficient [$C_I^{(n)}$].
Thus, the initiator approximation considers only the importance of the parent determinant, while as already mentioned selection performed by the PCI considers both the importance of parent determinants and the coupling between parent and spawned determinants.

\subsection{Sources of errors in the PCI method}
When compared to FCI, the PCI method introduces two types of error.
The first, the truncation error, is connected to the use of a subset of the full Hilbert space of determinants, and also affects selected CI methods.
Note that the truncation error does not affect methods like FCIQMC, which in principle can sample the entire Hilbert space.
The second type of error, the path filtering error, arises from approximating the action of the generator onto a state vector via Eqs.~\eqref{eq:eta_amplitude} and \eqref{eq:coef-propagation}.
The path filtering error may be viewed as arising from the diagonalization of an approximate Hamiltonian ($\tilde{H}$), which results from the path filtering procedure:
\begin{equation}
\tilde{H}^{(n)}_{IJ} =
\begin{cases}
 H_{IJ} &\text{ if } A^{(n)}_{IJ}(\eta) \neq 0\\
 0   &\text{ if } A^{(n)}_{IJ}(\eta) = 0
 \end{cases}
\end{equation}
Obviously, $\tilde{H}^{(n)}$ depends on the current wave function, and it is not guaranteed to be symmetric since in general $A^{(n)}_{IJ}(\eta) \neq A^{(n)}_{JI}(\eta)$.
In the PCI, the path filtering error arises from the fact that the wave function coefficient vector is the right eigenvector of $\tilde{H}^{(n)}$, which differs from the eigenvector of the full Hamiltonian in the subset $\detset{n}$.
Note, that the initiator approximation used in the FCIQMC approach is a form of path filtering, and consequently, it introduces a source of error analogous to the path-filtering error.

\section{Implementation}
\label{sec:range}
\subsection{The PCI algorithm}
The determinant selection procedure implemented via path filtering may be combined with the repeated application of the generator to obtain an approximate representation of the ground state wave function.
In the case of the linear generator the resulting PCI algorithm consists of the following steps:
\begin{enumerate}
\item Trial wave function generation.
The PCI procedure starts by selecting a trial wave function $\Omega^{(0)}$ to which corresponds the determinant space $\detset{0}$.  Although a convenient choice for the initial trial wave function $\Omega^{(0)}$ is the Hartree--Fock determinant $\Phi_{\rm HF}$, a CI with selected single and doubles out of $\Phi_{\rm HF}$ yields faster convergence to the ground state.

\item Range estimation.  The expectation value of the Hamiltonian with respect to the initial guess, $\bra{\Omega^{(0)}}\hat{H}\ket{\Omega^{(0)}}$ is used to estimate an upper bound to the ground state energy $E_0$. 
To estimate an upper bound to the energy of the highest excited state $E_{N-1}$, we employ Gershgorin's circle theorem.
%This theorem asserts that the eigenvalues of $\hat{H}$ represented in the FCI basis are contained in the circles, $D(H_{II},R_I)$, centered around $H_{II}$ with radius
%\begin{equation}
%R_I = \sum_{J (\neq I)} |H_{IJ}|
%\end{equation}
Accordingly, we approximate the upper bound to the eigenvalues of $\hat{H}$ as the sum of the diagonal element with the highest energy ($\bra{\Phi_{N-1}} \hat{H} \ket{\Phi_{N-1}}$) plus the sum of the absolute values of the off-diagonal matrix elements that couple $\ket{\Phi_{N-1}}$ to other determinants:
%\begin{equation}
%\begin{split}
%E_{N-1} \leq & \bra{\Phi_{N-1}} \hat{H} \ket{\Phi_{N-1}}\\
%&+\sum_{i \in \Phi_{N-1}}\sum_{a \notin \Phi_{N-1}} |f_{ia}|\\
%&+\frac{1}{4}\sum_{i,j \in \Phi_{N-1}}\sum_{a,b \notin \Phi_{N-1}} |\aphystei{ij}{ab}|
%\end{split}
%\end{equation}
%where the indices $i$,$j$ and $a$,$b$ are the spin orbitals that are occupied and unoccupied in $\Phi_{N-1}$, respectively, while $f_{ia}$ is the Fock operator between $i$ and $a$. The parameters for projector generator are then determined by the range of Hamiltonian.

\begin{equation} \label{eq:Gershgorin}
\tilde{E}_{N-1} = \bra{\Phi_{N-1}} \hat{H} \ket{\Phi_{N-1}}+\sum_{I}^{N-2} |\bra{\Phi_{N-1}} \hat{H} \ket{\Phi_{I}}|
\end{equation}
This estimate is not guaranteed to be a strict upper bound to $E_{N-1}$ since it is possible that other Gershgorin circles might enclose energy ranges higher than the value of Eq.~\eqref{eq:Gershgorin}.

\item Propagation step.
At step $n$, for each determinant $\Phi_J \in \detset{n}$ loop over all the singly and doubly excited determinants $\Phi_I$:
\begin{equation}
\Phi_I \in 
\{
\cop{a}\aop{i}\Phi_J,
\cop{a}\cop{b}\aop{j}\aop{i}\Phi_J \}
\end{equation}
where the indices $i,j$ ($a,b$) label occupied (virtual) orbitals of $\Phi_J$.
For each determinant $\Phi_I$, compute the thresholded spawning amplitude [$A_{IJ}^{(n+1)}(\eta)$] according to Eq.~\eqref{eq:eta_amplitude} and add it to the wave function coefficient $\tilde{C}_I^{(n)}$:
\begin{equation}
\tilde{C}_I^{(n+1)} \leftarrow  \tilde{C}_I^{(n+1)} + A_{IJ}^{(n+1)}(\eta)
\end{equation}
Since the propagation step can be performed independently for each of the determinant in $\detset{n}$, this section of the PCI algorithm may be easily parallelized by distributing the evaluation of $\tilde{C}_I^{(n+1)}$ over multiple threads/instances.

\item Normalization.  The wave function at step $n+1$ is normalized according to 
\begin{equation}
C_I^{(n+1)} = \frac{\tilde{C}_I^{(n+1)}}{\|\tilde{C}^{(n+1)}\|_2} \quad \forall \Phi_I \in \detset{n+1}
\end{equation}
where $\|\tilde{C}^{(n+1)}\|_2$ is the 2-norm of the vector $\tilde{C}^{(n+1)}$.

\item Energy evaluation.
The updated wave function coefficients are used to estimate the energy using two approaches.
The first is the variational estimator [$E^{(n)}_{\rm var}$], which is given by the expectation value of the PCI wave function:
\begin{equation} \label{eq:e_var}
E^{(n)}_{\rm var} =\bra{\Omega^{(n)}}\hat{H} \ket{\Omega^{(n)}}= \sum_{IJ} C^{(n)}_I H_{IJ} C^{(n)}_J
\end{equation}
The evaluation of $E_{\rm var}$ scales as $O^2V^2 N_{\rm det}$, where $N_{\rm det}$ is the number of determinants in $S^{(n)}$, therefore it has a computational cost comparable to that of applying $\hat{H}$ without path filtering.
% computationally more demanding than computing $E_{\rm proj}$.
Nevertheless, $E_{\rm var}$ is an upper bound to the exact ground state energy and the error is quadratic in the error of the wave function.
To speed up the evaluation of $E_{\rm var}$ during the iterative procedure we apply numerical screening to the vector $C_I^{(n)}$.

We also compute the energy via the projective estimator [$E_{\rm proj}^{(n)}$], defined as:
\begin{equation} \label{eq:e_proj}
E^{(n)}_{\rm proj}(J) = H_{JJ} + \sum_{I (\neq J)} H_{IJ}  \frac{C^{(n)}_I}{C^{(n)}_J}
\end{equation}
where $H_{IJ} = \bra{\Phi_J}\hat{H} \ket{\Phi_I}$ and $\Phi_J$ is chosen to be the determinant with the largest contribution to the wave function, that is, $J = \arg\max_{I}|C^{(n)}_I|$.
$E_{\rm proj}$ may be evaluate with a cost proportional to $O^2V^2$, where $O$ and $V$ are the number of occupied and virtual orbitals, respectively.
However, the projective estimator is not variational and its error is linear in the wave function error.
Consequently, the projective estimator is only used to monitor the convergence of the PCI algorithm.

\item Convergence check.  Evaluate the approximate energy gradient:
\begin{equation} \label{eq:energy_gradient}
\delta E^{(n+1)} =  \frac{1}{\gamma} (E^{(n+1)} - E^{(n)}),
\end{equation}
where $\gamma$ is the convergence factor of the projector generator.
If $|\delta E^{(n+1)}|$ is larger then the convergence threshold increase $n$ by one and go to Step 2.
Otherwise, the computation is converged and the final variational energy is evaluated including all contributions from the truncated CI space $\detset{n+1}$.
\end{enumerate}
The PCI algorithm is implemented in \textsc{Forte}, a suite of multireference electronic structure methods\cite{FORTE2016} written as a plugin to the open-source quantum chemistry package \textsc{Psi4}.\cite{turney2012psi4}

\section{Results}
Unless otherwise noted, all the PCI calculations are performed with the $5^{\rm th}$-order wall-Chebyshev generator.
PCI results obtained with a spawning threshold equal to $\eta$ are labeled as PCI($\eta$).
Preliminary computations showed that the variational estimator [Eq.~\eqref{eq:e_var}] yields energy errors that are consistently one order of magnitude smaller than those from than the projective estimator [Eq.~\eqref{eq:e_proj}].
Consequently, all results presented in this work are based on the variational energy estimator.

\subsection{N$_2$}

\begin{figure}[t!]
\includegraphics[width=3.375in]{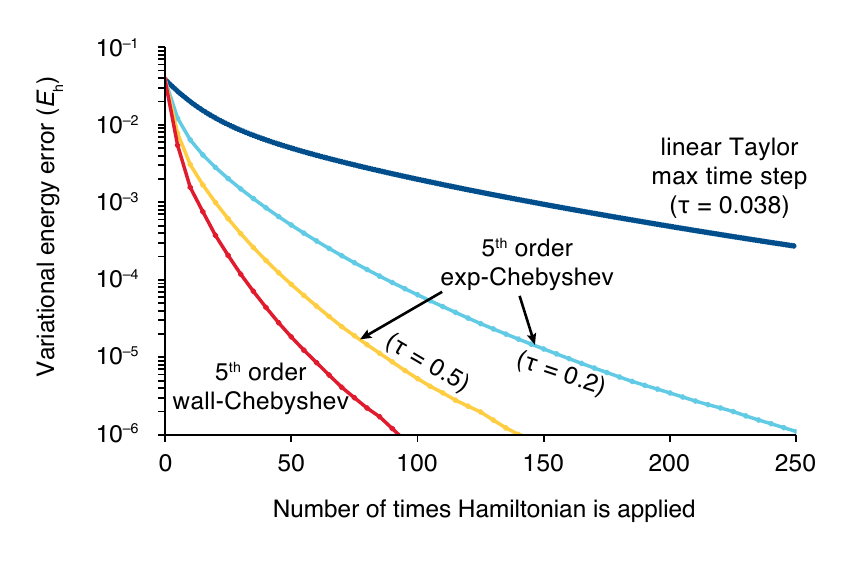}
\caption{Ground state of N$_2$ at the equilibrium geometry ($r=2.118$ bohr) computed with the PCI using a spawning threshold $\eta=1\e{-5}$ and various projector generators.
Difference between the variational energy at a given iteration and the converged energy as a function of the number of times the Hamiltonian is applied.
 All computations used canonical Hartree--Fock orbitals and the cc-pVDZ basis set. The 1s-like orbitals of nitrogen were excluded from computations of the correlation energy.}
\label{fig:convergence}
\end{figure}

\setlength{\tabcolsep}{0pt}
\begin{table*}[t!]
\begin{threeparttable}
\footnotesize
%\tiny
  \caption{
  Comparison of the ground state energy of N$_2$ computed with the PCI and several wave function approaches using the cc-pVDZ basis at  equilibrium and stretched bond lengths ($r$=2.118 and 4.2 bohr).  
$\Delta E$ is the energy error with respect to FCI computed with the variational estimate.
$N_{\rm par}$ is the number of variational parameters, with values in parentheses indicating the number of perturbative parameters.
$\rm NPE$ is the non-parallelism error defined by the difference of energy errors between stretched and equilibrium geometries.
All PCI computations use canonical restricted Hartree--Fock orbitals. The nitrogen 1s-like orbitals were frozen in all computations of the correlation energies.}
   \label{tab:n2}
   { \def\arraystretch{1.2}\tabcolsep=0pt    
     \begin{tabular*}{6.5in}{@{\extracolsep{\fill} } lrrrrr}
       
       \hline
       
       \hline
       \multirow{2}{*}{Method}  & \multicolumn{2}{c}{$r=2.118$ bohr} & \multicolumn{2}{c}{$r=4.2$ bohr}& \multirow{2}{*}{NPE/m\Eh}\\
       \cline{2-3} \cline{4-5}
%       & \multicolumn{3}{c}{} & \multicolumn{2}{c}{MP2} & \multicolumn{2}{c}{RHF} & \multicolumn{2}{c}{MP2} & RHF & MP2 \\
         & $N_{\rm par}$ & $\Delta E$/\Eh & $N_{\rm par}$ & $\Delta E$/\Eh \\
       \hline
       MP2 & (2,090) & 1.56\e{-2 } & (2,090) & $-$3.07\e{-1 } & $-$322.275 \\
       CISD & 2,090 & 3.65\e{-2 } & 2,090 & 2.64\e{-1 } & 227.215 \\
       CISDT & 60,842 & 2.59\e{-2 }& 60,842 & 2.41\e{-1 } &  215.173 \\
       CISDTQ & 969,718 & 2.31\e{-3 } & 969,718 & 5.72\e{-2 }  & 54.855 \\
       CCSD & 2,090 & 1.45\e{-2 } & 2,090 & 4.07\e{-2 } & 26.234 \\
       CCSD(T) & (58,752) & 1.87\e{-3 } & (58,752) & $-$1.65\e{-1 } & $-$166.876 \\
       \\
       MRCISD\tnote{a} & $\cdots$ & 6.64\e{-3} & $\cdots$ &6.91\e{-3} & 0.259\\
       MRCCSD\tnote{a} & $\cdots$ & 1.52\e{-3} & $\cdots$ &2.25\e{-3} & 0.732\\
       \\
PCI(1\e{-3}) & 12,393 & 2.45\e{-2} & 30,379 & 2.63\e{-2} & 1.816  \\
PCI(1\e{-4}) & 292,858 & 4.87\e{-3} & 573,665 & 1.06\e{-2} & 5.709 \\
PCI(5\e{-5}) & 532,728 & 3.08\e{-3} & 1,108,882 & 6.03\e{-3} & 2.952  \\
PCI(2\e{-5}) & 1,264,528 & 1.57\e{-3} & 2,628,056 & 2.25\e{-3}& 0.682 \\
PCI(1\e{-5}) & 2,703,218 & 8.76\e{-4} & 4,630,411 & 9.69\e{-4} & 0.093 \\
PCI(1\e{-6}) & 22,855,011 & 7.30\e{-5} & 32,900,610 & 8.82\e{-5}  & 0.015  \\
\\
PCI(1\e{-3})+diag\tnote{b} & 12,393 & 1.32\e{-2} & 30,379 & 1.55\e{-2} & 2.276\\
PCI(1\e{-4})+diag\tnote{b}  & 292,858 & 1.51\e{-3} & 573,665 & 2.68\e{-3} & 1.171\\
PCI(1\e{-5})+diag\tnote{b}  & 2,703,218 & 1.68\e{-4} & 4,630,411 & 1.82\e{-4} & 0.014 \\
PCI(1\e{-6})+diag\tnote{b}  & 22,855,011 & 8.39\e{-6} & 32,900,610 & 9.12\e{-6} & 0.001\\
       \\
       FCI         & 540,924,024 &                 & 540,924,024 &                       &  \\

       \hline
       
       \hline
      \end{tabular*}
      }
    \begin{tablenotes}
      \item[a] MRCISD and MRCCSD data based on a CASSCF(6\textit{e},6\textit{o}) reference wave function were taken from Ref.~\citenum{Chan2004State}.
      \item[b] The PCI+diag energies are computed by diagonalizing the Hamiltonian in the space of determinants obtained from a converged PCI computation.
      \end{tablenotes}
\end{threeparttable}
\end{table*}

To investigate the properties of the PCI approach we report computations of the ground state energy of the nitrogen molecule using the cc-pVDZ basis set\cite{Dunning1989Gaussian} and freezing the 1s core orbitals.
We discuss both the equilibrium ($r = 2.118$~bohr) and stretched ($r = 4.2$~bohr) geometries of N$_2$.

Figure~\ref{fig:convergence} illustrates the difference in efficiency between various generator at the equilibrium geometry.
%the linear generator, the fifth-order exponential-Chebyshev generator, and the fifth-order wall-Chebyshev generators at the equilibrium geometry.
To facilitate the comparison among the various generators, we plot the energy error with respect to the number of times $\hat{H}$ is applied to a state vector using a spawning threshold equal to $10^{-5}$.
For the linear generator we select $\tau = 1/R  = 0.038$ \invEh, the largest value of $\tau$ compatible with the spectral range of $\hat{H}$ [see Eq.~\eqref{eq:gamma_lin}].
As illustrated in Figure~\ref{fig:convergence}, the linear generator shows very slow convergence.
After 250 steps, the total error is still larger than $10^{-4}$ \Eh.  
%We 
% obtained from  () and the first-order wall-Chebyshev generator (which corresponds to a value of $\tau=0.025$ \invEh).
%Although the latter corresponds to a linear generator with smaller convergence factor, it nevertheless requires less iterations than the linear generator with the largest allowed $\tau$.
%The different performance of these two generators illustrates the importance of using a generator with a large convergence factor, which at the same time can efficiently reduce the coefficients of all excited states.
Projectors based on the exp-Chebyshev generators allow to use larger values of $\tau$ and converge more readily.
For example, with $\tau = 0.5$ \invEh, the fifth-order exp-Chebyshev projector requires 150 applications of $\hat{H}$ to achieve an error less than $10^{-6}$ \Eh.
The fifth-order wall-Chebyshev generator (which correspond to the limit $\tau\rightarrow\infty$) is more efficient than the exp-Chebyshev generators as it can achieve the same level of accuracy with less than 100 applications of $\hat{H}$.

%\begin{figure}[t!]
%\includegraphics[width=3.375in]{figure_eta.pdf}
%\caption{Variational energy computed with PCI using different spawning thresholds. All computations used canonical Hartree--Fock orbitals and the cc-pVDZ basis set. The 1s-like orbitals of nitrogen were excluded from computations of the correlation energy. Number of determinants involved in PCI calculation using different spawning thresholds. All computations used canonical Hartree--Fock orbitals and the cc-pVDZ basis set. The 1s-like orbitals of nitrogen were excluded from computations of the correlation energy.}
%\label{fig:eta}
%\end{figure}
 
Next, we study the accuracy of the PCI as a function of the spawning threshold ($\eta$) and compare it to a selection of single-reference and multireference methods.
Table~\ref{tab:n2} reports a comparison of the total error with respect to FCI for the variational energy estimator [Eq.~\eqref{eq:e_var}].
Additionally, Table~\ref{tab:n2} reports energies for N$_2$ computed using second-order M{\o}ller--Plessett (MP2) perturbation theory, truncated CI with up to quadruple excitations (CISD--CISDTQ), coupled cluster with singles and doubles (CCSD), CCSD with perturbative triples corrections [CCSD(T)], uncontracted multireference CISD (MRCISD), and multireference CCSD (MRCCSD) based on a CASSCF(6\textit{e},6\textit{o}) reference.\cite{Chan2004State}
%Single-reference approaches are found to yield accurate energies (errors in the range 4\e{2}--1\e{3} \Eh) at the equilibrium bond lengths.

From Table~\ref{tab:n2} it can be seen that since the PCI wave function is not biased towards a reference determinant, it can efficiently capture both static and dynamic electron correlation and provide an accurate description of N$_2$ at both  equilibrium and stretched geometries.
For example, even with a large spawning threshold ($\eta =$ 1\e{-3}) the PCI yields a non-parallelism error (NPE, defined as the difference in energy error between the equilibrium and stretched geometries) that is of the order of a few m\Eh.
In contrast, single-reference approaches give NPEs that range from $-322$ to $+227$ m\Eh.

The accuracy of the PCI is effectively tuned by the spawning threshold and can be chosen to match or go beyond that of MRCI and MRCC.
For example, for $\eta$ = 2\e{-5}, the NPE is equal to 0.37 \kcal, which is within chemical accuracy (defined as an error less than 1 \kcal).
At the equilibrium geometry the PCI wave function has 1,264,528 determinants, with the Hartree--Fock determinant having a coefficient equal to 0.94.
At the stretched geometry, when the  coefficient of the Hartree--Fock determinant is only 0.46, this number increases to 2,628,056 determinants to accommodate the multideterminantal character of the wave function.
Note that at both geometries the PCI(2\e{-5}) wave function uses less than 0.5\% of the FCI space determinants.

In order to illustrate the importance of the truncation and path-filtering errors, in Table~\ref{tab:n2} we report energies obtained by diagonalizing the Hamiltonian in the PCI determinant space (indicates as PCI+diag).
These energies are more accurate than the corresponding PCI values.
For example, with $\eta = $ 1\e{-4}, the NPE for the PCI and PCI+diag are 5.7 and 1.2 m\Eh, respectively.
The difference between the energy from FCI and PCI+diag represents the truncation error, while the gap between the PCI and PCI+diag energies is the path-filtering error.
For large spawning thresholds (e.g. $\eta$ = 1\e{-3}) the truncation and path-filtering errors contribute equally to the total error. However, as the spawning threshold decreases, path-filtering becomes the dominant source of error.
For example, when the spawning threshold is equal to 1\e{-6}, the path-filtering error contributes to 90\% of the total error. In this case, the diagonalization of the PCI space yields energies within 10 $\mu$\Eh from FCI values, while the nonparallelism error is about 1 $\mu$\Eh.

To give an idea of the computational cost of the PCI, we note that the N$_2$ computations at equilibrium geometry with $\eta = 10^{-6}$ ran in 3 hours on 16 threads on a single node (on two Intel Xeon E5-2650 v2 processors) and took 16 iterations to finish.
%In this case, the PCI wave function has about 22 million determinants, a number that is similar to the largest selected CI calculations reported in literature.\cite{stampfuss2005improved}
The corresponding computation at the stretched geometry ran in 44 hours and took 127 iterations.  In this example the the wave function contains 33 million determinants and convergence is slowed by the small energy gap between the ground and first excited state.

\subsection{C$_2$}

\begin{table*}[t!]
\footnotesize
\begin{threeparttable}
  \caption{Comparison of the ground state energy of C$_2$ calculated with the PCI and several wave function approaches using the cc-pV$X$Z basis set ($X$ = D,T,Q).  All PCI computations use MP2 natural orbitals. The carbon 1s-like orbitals were frozen in all calculations. $N_{\rm par}$ indicates the number of variational parameters, with values in parentheses indicating perturbative parameters.  All results are shifted by +75 \Eh.}
   \label{tab:c2}
%   \newcolumntype{d}[1]{D{.}{.}{#1}}
    { \def\arraystretch{1.2}\tabcolsep=0pt    
     \begin{tabular*}{6.5in}{@{\extracolsep{\fill} }lcccccc}
       \hline
       
       \hline
       \multirow{2}{*}{Method} & 
       \multicolumn{2}{c}{cc-pVDZ (8\textit{e}, 26\textit{o})} &
       \multicolumn{2}{c}{cc-pVTZ (8\textit{e}, 58\textit{o})} & \multicolumn{2}{c}{cc-pVQZ (8\textit{e}, 108\textit{o})}
       \\ \cline{2-3} \cline{4-5} \cline{6-7}
        & $(E+75)$/\Eh & $N_{\rm par}$ & $(E+75)$/\Eh & $N_{\rm par}$ & $(E+75)$/\Eh & $N_{\rm par}$\\
       \hline
       MP2 & $-$0.697 678 & (1.43\e{3}) & $-$0.756 562 & (8.35\e{3}) & $-$0.777 234 & (3.05\e{4}) \\
       CISD & $-$0.663 765 & 1.43\e{3} & $-$0.711 300 & 8.35\e{3} & $-$0.726 551 & 3.05\e{4} \\
       CISDT & $-$0.682 929 & 3.34\e{4} & $-$0.733 939 & 4.96\e{5} & $-$0.749 947 & 3.55\e{6}\\
       CISDTQ & $-$0.721 845 & 4.11\e{5} & $-$0.777 182 & 1.51\e{7}  & $-$0.794 504 & 2.09\e{8}\\
       CCSD & $-$0.699 132 & 1.43\e{3} & $-$0.749 551 & 8.35\e{3} & $-$0.765 696 & 3.05\e{4} \\
       CCSD(T) &$-$0.726 697 & (3.20\e{4}) & $-$0.783 070 & (4.88\e{5}) & $-$0.800 807 & (3.52\e{6}) \\
       \\
       PCI(1\e{-4}) & $-$0.725 914 & 1.58\e{5} & $-$0.779 959 & 5.67\e{5} & $-$0.796 216 & 1.00\e{6}\\
       PCI(5\e{-5}) & $-$0.727 131 & 3.09\e{5} & $-$0.781 984 & 1.27\e{6} & $-$0.798 720 & 2.40\e{6}\\
       PCI(1\e{-5}) & $-$0.728 292 & 1.22\e{6} & $-$0.784 133 & 7.45\e{6} & $-$0.801 450 & 1.67\e{7}\\
       PCI(5\e{-6}) & $-$0.728 439 & 2.03\e{6} & $-$0.784 561 & 1.50\e{7} & $-$0.801 973 & 3.65\e{7}\\
       PCI(1\e{-6}) & $-$0.728 541 & 5.56\e{6} & $-$0.784 961 & 6.79\e{7} & $-$0.802 513 & 1.99\e{8} \\
       PCI(extrapol.)\tnote{a} & $-$0.728 565 &  & $-$0.785 069 & & $-$0.802 665 & \\
       \\
       DMRG\tnote{b,c} & $-$0.728 556 & 5.2\e{5} & $-$0.785 054 & 1.2\e{7} & $-$0.802 671& 7.0\e{7} \\
       DMRG\tnote{d} & $\cdots$ & $\cdots$ & $\cdots$ & $\cdots$ & $-$0.802 69$\;\;$& $\cdots$ \\%[3.9\e{7}] \\       
       i-FCIQMC\tnote{e,f} & $-$0.728 78$\;\;$ & 4.2\e{6} & $-$0.784 93$\;\;$ & 6.3\e{6} & $-$0.802 51$\;\;$ & 3.0\e{7}\\
       i-SFCIQMC\tnote{g,f} & $\cdots$ &$\cdots$ & $\cdots$ & $\cdots$ & $-$0.802 575 & 1.6\e{7}\\

       FCI\tnote{h} & $-$0.728 556 & 2.79\e{7} & $\cdots$ & 2.25\e{10} & $\cdots$ & 3.59\e{12}\\
       \hline
       
       \hline
      \end{tabular*}
      \begin{tablenotes}
      \item[a] Extrapolated PCI values obtained from a quadratic fitting of the results with $\eta$ = $10^{-5}$, $5 \times 10^{-6}$, and $10^{-6}$.   
      \item[b] DMRG data taken from Ref.~\citenum{Olivares-Amaya2015ab}.  Based on the genetic algorithm ordering and accurate to better than 0.01 m\Eh.
      \item[c] DMRG number of variational parameters were kindly provided by Guo and Chan \cite{Garnet:private2016} for computations with 946, 3234, and 6738 renormalized states using the DZ, TZ, and QZ basis sets, respectively.
      \item[d] DMRG data taken from Ref.~\citenum{sharma2015general}.  Based on the genetic algorithm ordering and accurate to better than 0.01 m\Eh.
      \item[e] Initiator FCIQMC (i-FCIQMC) data taken from Ref.~\citenum{Booth2011Breaking}.
      \item[f] For i-FCIQMC and i-SFCIQMC the column labeled $N_{\rm par}$ reports the total number of walkers.
      \item[g] Initiator semi-stochastic FCIQMC (i-SFCIQMC) data taken from Ref.~\citenum{blunt2015excited-state}.      
      \item[h] The number of FCI determinants for the triple- and quadruple-$\zeta$ basis sets was estimated as $\binom{N_{\rm orb}}{N_{\rm el}/2}^2/N_{\rm irrep}$, where $N_{\rm orb}$, $N_{\rm el}$, and $N_{\rm irrep}$ are the number of orbitals, electrons, and irreps, respectively.      
      \end{tablenotes}
}
\end{threeparttable}
\end{table*}

To study the performance of PCI on larger basis sets we computed the ground state energy of C$_2$ at the equilibrium geometry using basis sets that range from double- to quadruple-$\zeta$ quality.
Table~\ref{tab:c2} collects PCI results obtained using MP2 natural orbitals, together with truncated configuration interaction, coupled cluster, DMRG, and i-FCIQMC results.
When possible, computations were reported for the  first three basis sets of the cc-pV$X$Z series ($X$=D,T,Q, also abbreviated as XZ in the following discussion).\cite{Dunning1989Gaussian,kendall1992electron}
For the TZ and QZ basis sets the FCI energy cannot be computed, and we take DMRG results from Ref.~\citenum{Olivares-Amaya2015ab} as a reference.
PCI($\eta$) energies are extrapolated to zero spawning threshold by fitting results with $\eta =$ 1\e{-5}, 5\e{-6}, and 1\e{-6} to a quadratic function.

Table~\ref{tab:c2} illustrates how the PCI energy may be systematically converged to the reference FCI/DMRG energy with control over the absolute energy error.  For example, with a spawning threshold equal to $10^{-5}$, for all basis sets the PCI energy is within 1.3 m\Eh from the DMRG energy.
%0.000258 DZ
%0.000921 TZ
%0.001219 QZ
While with a spawning threshold equal to $10^{-6}$, the error is further reduced to less than 0.2 m\Eh in all cases.
% with a spawning threshold equal   in different basis sets, as shown in Tab.~\ref{tab:c2} column-wise, the energy consistently converges to the reference energy as spawning threshold decreases. Sub m\Eh accuracy is easily achieved with only a fraction of determinants.

When compared to other methods, the cheapest PCI calculations ($\eta = 10^{-4}$) shown in Tab. \ref{tab:c2} are found to be already more accurate than truncated CI methods up to quadruple excitation and CCSD.
Moreover, the PCI selects the most important determinants efficiently and therefore shows a more favorable accuracy/(number of parameters) ratio.
For example, the cc-pVQZ PCI($\eta = 10^{-4}$) wave function has about one million determinants, but yields an energy that is more accurate than that of CISDT (3 million determinants) and CISDTQ (200 million determinants). We note that PCI results surpass the accuracy of the CCSD(T) method with a spawning threshold of 1\e{-5}.

The PCI shows a favorable scaling with respect to the size of basis set.
When the basis set is enlarged from DZ to QZ, the number of orbitals involved in calculation grow from 26 to 108 and the corresponding FCI space increased ca. $10^5$ folds.
The corresponding growth of PCI determinants with respect to the number of virtual orbitals ($n_{\rm virt}$) is found to be linear, with increase of only 14 and 36 times when $\eta=$1\e{-5} and $1\e{-6}$, respectively.
In comparison, truncated CI and CC schemes scale as $n_{\rm virt}^2$, $n_{\rm virt}^3$, and $n_{\rm virt}^4$ for the SD, SDT, and SDTQ truncation schemes, respectively.  Consequently, the cost of these computations grows by a factor 21, 106, and 509 when going from the DZ to the QZ basis set.
FCIQMC also shows very good scaling with respect to virtual orbitals, with an increase of only about 7 times the number of walkers.
In the case of DMRG, assuming that the number of renormalized states ($M$) required to obtain a given level of accuracy scales as $n_{\rm occ} n_{\rm vir}$,\cite{Olivares-Amaya2015ab} then the number of variational parameters scales as $n_{\rm orb} M^2 \approx n_{\rm vir}^3$.

We would like to point out that the QZ PCI calculation with spawning threshold 1\e{-6} (200 million determinants) ran on a single node.
This computation is two orders of magnitude larger than the largest selected CI calculations reported in the literature (4 million determinants),\cite{stampfuss2005improved} which was performed with a parallel algorithm on a distributed memory architecture with 32--256 nodes.
As a comparison, typical FCIQMC computations may employ up to 2--7 billion walkers.\cite{Booth2009Fermion}

%In each column of table \ref{tab:c2}, as spawning threshold approaches zero, the computed results approach benchmark values. Besides, the energy errors are also controlled by spawning threshold, though the error is larger for larger basis set computation because relatively more determinants are excluded from selected space and increased the energy error.
%
%The successfully computed energy with cc--pVQZ basis set demonstrates that PIFCI is capable of compute system with large basis set. A feature for PIFCI is that the computational cost does not increase dramatically when we switch to larger basis set or do all electron computation rather than frozen core computation. For example, in all electron calculation with spawning threshold of 5\e{-5}, the number of determinants in CI space is only 16 times larger, rather than exponentially grow. Switching from frozen core computation to all election also only added around 80\% more determinants in CI space. Thus, PIFCI provides accurate results with bounded resources.

\subsection{Size consistency and molecular orbital comparison}
\begin{table*}[t!]
\begin{threeparttable}
\footnotesize
  \caption{Analysis of the size consistency error ($\Delta E$) of truncated CI methods and the PCI for the (Be--He)$_2$ system.
%   as a function of the spawning threshold and the type of molecular orbitals for the (Be--He)$_2$ system.  
   All results used the cc-pVDZ and STO-3G basis sets for Be and He, respectively.  The Be--He bond distance in the monomer is equal to 2.5 \AA{}.  The column labeled $N_{\rm det}$ reports the size of each CI space.}
   \label{tab:size}
    {  \def\arraystretch{1.2}\tabcolsep=0pt    
     \begin{tabular*}{6.5in}{@{\extracolsep{\fill} }lrrrrr}
       \hline
       
       \hline
       & \multicolumn{2}{c}{Be--He (6\textit{e},15\textit{o}) } & \multicolumn{2}{c}{He--Be $\cdots$ Be--He (12\textit{e},30\textit{o})} \\\cline{2-3} \cline{4-5}
       Method & Energy/\Eh & $N_{\rm det}$ & Energy/\Eh  & $N_{\rm det}$ & $\Delta E$/m\Eh \\
       \hline
%       \multicolumn{6}{c}{Be/Be$_2$ cc-pVDZ (4\textit{e},14\textit{o})/(8\textit{e},28\textit{o})} \\[3pt]
%       RHF & $-$14.572 338 & 1 & $-$29.144 675 & 1 & 0.000\\
%       FCI & $-$14.617 410 & 1,093 & $-$29.234 819 & 52,407,353  & 0.000\\
%       CISD & $-$14.617 356 & 121 & $-$29.227 666 & 1,929  & 7.046\\
%       CISDT & $-$14.617 393 & 529 & $-$29.227 759 & 45,601  & 7.027\\
%       CISDTQ & $-$14.617 410 & 1,093 & $-$29.234 808 & 589,397  & 0.012\\
%       \multicolumn{6}{c}{Canonical Hartree--Fock orbitals} \\
%       PCI : 1\e{-4} & $-$14.617 404 & 370 & $-$29.234 519 & 19,216 & 0.290 \\
%       PCI : 1\e{-5} & $-$14.617 409 & 685 & $-$29.234 798 & 89,641 & 0.021 \\
%       PCI : 1\e{-6} & $-$14.617 410 & 1,007 & $-$29.234 818 & 274,501 & 0.001 \\
%       \multicolumn{6}{c}{MP2 natural orbitals} \\
%       PCI : 1\e{-4} & $-$14.617 408 & 302 & $-$29.234 760 & 14,235 & 0.056 \\
%       PCI : 1\e{-5} & $-$14.617 409 & 731 & $-$29.234 815 & 52,933 & 0.004 \\
%       PCI : 1\e{-6} & $-$14.617 410 & 1,013 & $-$29.234 818 & 235,160 & 0.001 \\
%%       PCI : 1\e{-4} & $-$14.617 408 & 306 & $-$29.234 766 & 14,235 & 0.050 \\
%%       PCI : 1\e{-5} & $-$14.617 409 & 731 & $-$29.234 814 & 49,935 & 0.004 \\
%%%       PCI : 5\e{-6} & $-$14.617 409 & 857 & $-$29.234 818 & 88 242 & 0.000 \\
%%       PCI : 1\e{-6} & $-$14.617 410 & 1,007 & $-$29.234 819 & 220,582  & 0.000\\
%       \hline
%       \multicolumn{6}{c}{BeHe/[BeHe]$_2$ Be: cc-pVDZ, He: STO-3G (6\textit{e},15\textit{o})/(12\textit{e},30\textit{o})} \\[3pt]
       RHF & $-$17.374 136 & 1 & $-$34.748 272 & 1 & 0.000\\
       FCI  & $-$17.420 556 & 51,853 & $-$34.841 113 & 4.41\e{10}  & 0.000\\ 
       CISD  & $-$17.420 420 & 523 & $-$34.833 525 & 4,405 & 7.316 \\
       CISDT  & $-$17.420 484 & 4,257 & $-$34.833 664 & 170,685 & 7.305 \\
       CISDTQ  & $-$17.420 556 & 17,973 & $-$34.841 084 & 3,833,121 & 0.029 \\
       \\
       \multicolumn{6}{c}{Delocalized canonical Hartree--Fock orbitals} \\
       PCI(1\e{-4})  & $-$17.420 537 & 1,424 & $-$34.840 544 & 34,164 & 0.529 \\
       PCI(1\e{-5})  & $-$17.420 556 & 5,311 & $-$34.841 066 & 255,342 & 0.045 \\
       PCI(1\e{-6})  & $-$17.420 556 & 15,465 & $-$34.841 108 & 1,558,745 & 0.005 \\
       \\
       \multicolumn{6}{c}{Delocalized MP2 natural orbitals} \\
       PCI(1\e{-4})  & $-$17.420 547 & 1,138 & $-$34.840 924 & 23,979 & 0.169 \\
       PCI(1\e{-5})  & $-$17.420 556 & 5,077 & $-$34.841 088 & 163,469 & 0.024 \\
       PCI(1\e{-6})  & $-$17.420 556 & 14,801 & $-$34.841 110 & 1,185,988 & 0.002 \\
       \\
       \multicolumn{6}{c}{Localized canonical Hartree--Fock orbitals} \\       
       PCI(1\e{-4})  & $-$17.420 537 & 1,424 & $-$34.840 981 & 9,746 & 0.092 \\
       PCI(1\e{-5})  & $-$17.420 556 & 5,311 & $-$34.841 104 & 60,740 & 0.007 \\
       PCI(1\e{-6})  & $-$17.420 556 & 15,465 & $-$34.841 112 & 337,662 & 0.001 \\
       \\
       \multicolumn{6}{c}{Localized MP2 natural orbitals} \\       
       PCI(1\e{-4})  & $-$17.420 547 & 1,138 & $-$34.841 064 & 5,910 & 0.029 \\
       PCI(1\e{-5})  & $-$17.420 556 & 5,077 & $-$34.841 109 & 41,580 & 0.003 \\
       PCI(1\e{-6})  & $-$17.420 556 & 14,801 & $-$34.841 113 & 247,364 & 0.000 \\
              
%       \hline
%       \multicolumn{6}{c}{Ne/Ne$_2$ cc-pVDZ (10,14)/(20,28)\tnote{a}} \\
%       RHF & $-$128.488 776 & 1 & $-$256.977 551 & 1  & 0.000\\
%       
%       FCI & $-$128.680 881 & 501,992 & $-$257.361 762 & 21.5\e{12}  & 0.000\\
%       CISD & $-$128.675 427 & 400 & $-$257.341 813 & 6,271  & 9.041\\
%       CISDTQ & $-$128.680 764 & 30,654 & $-$257.361 074 & 10,838,228  & 0.454\\
%       PCI : 1\e{-4} & $-$128.680 817 & 19,113 & $-$257.359 123 & 630,338 & 2.510 \\
%       PCI : 1\e{-5} & $-$128.680 877 & 71,893 & $-$257.361 178 & 4,937,107 & 0.577 \\
%       PCI : 1\e{-6} & $-$128.680 881 & 198,265 & $-$257.361 717 & 85,188,504 & 0.045 \\
%       PCI : 1\e{-4} & $-$128.680 817 & 19,014 & $-$257.359 127 & 631,330  & 2.507\\
%       PCI : 1\e{-5} & $-$128.680 877 & 68,468 & $-$257.361 178 & 4,932,351  & 0.576\\
%%       PCI : 5\e{-6} & $-$128.680 880 & 106 698 & $-$257.361 415 & 11 687 045 & 0.345\\
%       PCI : 1\e{-6} & $-$128.680 881 & 194,250 & $-$257.361 717 & 85,144,137 & 0.045\\
       
       \hline
       
       \hline
      \end{tabular*}
%      \begin{tablenotes}
%      \item[a] In Ne/Ne$_2$ section, data computed with canonical Hartree--Fock orbitals are shown for PCI calculation. MP2 natural orbital PCI data is almost identical, where the largest difference in energy is less than 0.004 m\Eh, and the difference in number of determinants ($N_{\rm det}$) is less than 1\%.
%      \end{tablenotes}
    }
\end{threeparttable}
\end{table*}

Lastly, we investigate the degree to which the PCI wave function lacks size consistency, and how different type of molecular orbitals affect its performance.
In our tests we have considered a monomer consisting of Be and He separated by 2.5 \AA{}.
In one set of computations two monomers are arranged in a $D_{\infty \rm h}$ geometry, so that the orbitals are delocalized over the two fragments.
Starting from the $D_{\infty \rm h}$ geometry, we obtained a $C_{\infty \rm v}$ structure in which the Be--He distances of the monomers are shortened and lengthened by $\pm 10^{-5}$ \AA{}, respectively.
This geometric change leads to localization of the molecular orbitals on one of the two monomers.
For both localized and delocalized molecular orbitals we considered canonical Hartree--Fock orbitals and MP2 natural orbitals.
%We have also considered the case in which the system 
%
%We consider two systems.  The first one is a pair of noninteracting beryllium atoms using canonical and MP2 natural orbitals.  The second systems is a pair of Be--He complexes that do not interact computed using delocalized canonical Hartree--Fock orbitals and localized orbitals.

Table~\ref{tab:size} reports the size consistency error ($\Delta E$) for a pair of noninteracting Be--He units as a function of the spawning threshold, where $\Delta E$ is defined as the energy difference between a non-interacting dimer (Be--He$\cdots$Be--He) and twice the energy of the monomer (Be--He):
\begin{equation}
\Delta E = E(\text{Be--He} \cdots \text{Be--He}) - 2 E(\text{Be--He}).
\end{equation}

As expected, the PCI energy is not size consistent, but a comparison with truncated CI methods shows that the corresponding error is significantly smaller in the case of PCI and can be effectively controlled via the spawning threshold.
In comparison to CISDTQ, which requires 3,833,121 determinants for the dimer computation, the PCI($10^{-6}$) with canonical orbitals requires only 1,558,745 determinants and leads to a size consistency error that is six times smaller.
When delocalized orbitals are used, going from canonical Hartree--Fock orbitals to MP2 natural orbitals leads to a reduction of the size consistency error of the PCI by a factor of ca. two.  At the same time, the use of MP2 natural orbitals also slightly reduces the number of determinants.

Upon localization of the orbitals we observe a significant reduction of the size consistency error and wave function size.
For example, localization of the canonical Hartree--Fock orbitals reduces the PCI($10^{-6}$) size consistency error and number of determinants by a factor of five.
The best performance is obtained by combining localization with MP2 natural orbitals.  In this case the overall size of the PCI wave function is reduced by a factor of 6 and the size consistency error is less than 0.001 m\Eh.
This comparison shows that the use of optimized orbitals can significantly reduce the computational cost of the PCI and the magnitude of the size consistency error.
%This observation 

%Note that in both cases the number of iterations before converge did not vary significantly.

%This comparison shows that orbital optimization before PCI iterations can reduce the computational cost significantly. It is because in both MP2 natural orbitals and localized orbitals, orbitals are optimized so that there are more overlap between important determinants. Consequently, the total number required for PCI calculation using same spawning threshold decrease with more accurate results.

%\subsection{Excited states}

%
% Conclusion
%
\section{Summary and conclusions}
In this paper, we introduced a general projector diagonalization approach and combined it with path filtering to create a novel projector configuration interaction (PCI) method.
Given an operator (matrix) $\hat{H}$, the projector diagonalization method seeks to obtain one of the eigenvectors of $\hat{H}$ via repeated application of the projector generator $g(\hat{H})$ onto a trial vector.
The projector generator is a matrix function designed to amplify the coefficient of one of the eigenvectors.
The focus of this work is on polynomial projector generators derived from the imaginary-time propagator, which project the trial wave function onto the ground electronic state.
To improve the performance of a Taylor expansion of the imaginary-time propagator, we discuss its approximation in terms of Chebyshev polynomials, and propose a new generator (wall-Chebyshev) with superior convergence properties. 

The PCI optimization process is formulated in terms of a dynamics in which each application of the projector generator is equivalent to a spawning process.
In this process, each determinant spawns singly and doubly excited determinants with a given spawning amplitude. 
In order to truncate the determinant space explored by the PCI algorithm, we consider a path filtering approach in which spawning amplitudes are truncated according to a user-provided \textit{spawning threshold} ($\eta$).
Path filtering applied at each step of the projector diagonalization controls the size of the PCI wave function and the accuracy of the energy by selecting important determinants that contribute the most to a given eigenstate.
In this respect, the PCI method is similar to selected CI, with the important difference that the former also approximates the diagonalization process to increase computational efficiency.

Since the PCI is not biased towards any reference determinants, it can describe dynamic and static electron correlation equally well.
This point is illustrated with computations of the energy of N$_2$ at equilibrium and stretched geometries.  As shown in Table~\ref{tab:n2}, the PCI($\eta = 2\e{-5}$) can predict the energy difference between these two geometries with a non-parallelism error equal to 0.682 m\Eh (0.43 \kcal) using only a small fraction of the Hilbert space of determinants (less than 0.5\%).
Additionally, we compare PCI with DMRG and FCIQMC using the carbon dimer as a challenging benchmark.
With a spawning threshold equal to $10^{-6}$, the PCI can match the accuracy of FCIQMC results, while PCI  extrapolated to the limit $\eta\rightarrow0$ yields total energies that are within 0.01 m\Eh of DMRG reference data.
%Given its favorable cost scaling with respect to the number of unoccupied orbitals, PCI appears to be competitive with DMRG and FCIQMC in computations with large basis sets.
We have also analyzed the extent of size consistency errors in PCI computations.
This error is effectively controlled by the spawning threshold and may be further reduced by using a localized basis.

One of the interesting features of the PCI algorithm is that it can be expressed as a series of update steps in which spawning amplitudes for different determinants can be computed independently with no communication.  
Moreover, the linear and wall-Chebyshev generators only require storage of two vectors of the size of the CI space.
These two features make the PCI amenable to computations with large CI spaces containing 10$^7$--10$^8$ determinants.
A parallel implementation of the PCI for distributed-memory machines would allow to further increase the size of the CI space.
%only require information on a parent determinant and coefficient. Thus, no communication is required between paths. In case linear or wall-Chebyshev generator is applied, computation only require two vectors at the size of current CI space to be stored. These are reasons why PCI is able to compute with a space of more than $10^8$ determinants. The capacity can be enhanced easily with massively parallel implementation.
%Given the fact that the PCI approach obtains the ground state simply by repeated application of a polynomial of the Hamiltonian to a trial wave function, we expect that it should be possible to create efficient parallel implementations of the PCI.
Both the PCI and FCIQMC use a sparse representation of the FCI wave function and present similar challenges when implemented on distributed memory architectures.  Therefore, the recent successful implementation of a parallel FCIQMC code\cite{Booth2014Linear} suggests that it should be possible to also produce an efficient parallel implementation of the PCI.

Currently, the PCI algorithm has been formulated to optimize the ground state.
However, several strategies may be explored to extend the PCI to electronic excited states.
One possibility is a state-specific approach in which excited states are optimized individually, while maintaining orthogonality with lower energy states.
An alternative is a multistate version of the PCI in which several states are optimized simultaneously.\cite{ten2013stochastic}
Since the convergence of the PCI depends on ratio of the first excitation energy and the spectral radius, $(E_1-E_0)/R$, a multistate version of the PCI would also be helpful to speed up convergence to the ground state in cases when this ratio is small.
Another interesting venue to explore is to use the PCI approach to target the density matrix at finite temperatures\cite{Blunt:2014kl,Malone:2015ds} or to compute approximate spectral densities of systems with a dense manifold of low-energy electronic states.\cite{Lin:2016fm}

%
% Appendices
%
\appendix

\section{Path filtering for polynomial generators}
\label{app:general_path_filtering}

In this appendix we report a generalization of the path filtering approach for polynomial generators $g(x)$ of order $m$ that have $m$ real roots ($s_i, i = 1,\ldots,m$).
In this case, $g(x)$ can be written as:
\begin{equation}
g(x) = \prod_{i=1}^m \frac{x-s_i}{E_0-s_i}
\end{equation}
and $g(\hat{H})\ket{\trial^{(n)}}$ may be computed by repeated application of a linear generator with modified shift to which path filtering is applied in all intermediate steps.
It it important to point out that the path-filtering algorithm presented here gives results that are consistent with those of the algorithm outlined in the paper, which applies only to linear generators. 

For convenience, we start by defining a series of normalized trial wave functions
\begin{equation}
\ket{\trial^{(n+1,i)}} =  \sum_{I \in \detset{n,i}} C^{(n+1,i)}_I \ket{\Phi_I}
\end{equation}
expanded over the space $\detset{n,i}$.
The coefficient vector for $i=0$ is given by:
\begin{equation}
C^{(n+1,0)}_I = C^{(n)}_I
\end{equation}
and spans the space $\detset{n+1,0} = \detset{n}$.

The coefficients $C^{(n+1,i)}_I$ for $i > 0$ are obtained from the unnormalized wave function coefficients [$\tilde{C}_I^{(n+1,i)}$]:
\begin{equation} \label{eq:gen_norm}
C^{(n+1,i)}_I = \frac{\tilde{C}_I^{(n+1,i)}}{\| \tilde{C}^{(n+1,i)}\|_2}
\end{equation}
which are obtained as the sum:
\begin{equation} \label{eq:general_update}
\tilde{C}_I^{(n+1,i)} =\sum_{\Phi_J \in \detset{n+1,i}}    A_{IJ}^{(n+1,i)}(\eta)
\end{equation}

The path-filtered spawning amplitudes [$A_{IJ}^{(n,i)}(\eta)$] that enter into Eq.~\eqref{eq:general_update} are obtained from the untruncated amplitudes [$A_{IJ}^{(n,i)}$]:
\begin{equation}
\label{eq:modified-amplitude}
A_{IJ}^{(n,i)} =
\bra{\Phi_I}\hat{H} - s_i \ket{\Phi_J} C_J^{(n,i-1)}
\end{equation}
and truncated according to:
\begin{equation} \label{eq:gen_amplitudes}
A_{IJ}^{(n,i)}(\eta) =
\begin{cases}
A_{II}^{(n,i)} &\text{ if } I = J\\
A_{IJ}^{(n,i)} \Theta(|A_{IJ}^{(n,i)}| - \eta) &\text{ if } I \neq J
\end{cases}
\end{equation}

The normalized coefficients are evaluated recursively for $i = 1,2,\ldots,m$ following Eqs.~\eqref{eq:gen_norm}--\eqref{eq:gen_amplitudes}.
Finally, the coefficients for the updated wave function are given by:
\begin{equation}
C_I^{(n+1)} = C_I^{(n+1,m)}.
\end{equation}
Note that to evaluate the application of factorizable generators with real zeros onto a trial vector requires storage of two vectors.  Thus, require the same amount of memory as the linear projector.

\section*{Acknowledgments}
The authors are grateful to Philip Shushkov and Michele Benzi for valuable discussions  concerning the theory of projectors.
The authors would also like to thank Sheng Guo and Garnet Chan for providing the number of wave function parameters for the DMRG computations on the carbon dimer.

This work was supported by start-up funds provided by Emory University.

\newpage
\footnotesize
%\bibliographystyle{achemso}
%\mciteErrorOnUnknownfalse
\bibliography{pci}

\end{document}